\documentclass[amsmath,amssymb]{revtex4}
\usepackage{graphicx}
\usepackage{dcolumn}
\usepackage{bm}

\begin{document}


\title{Quantum dynamical correlations:\\ 
Effective potential analytic continuation approach}

\author{Atsushi Horikoshi$^{1,2}$}
 \email{horikosi@cc.nara-wu.ac.jp}
\author{Kenichi Kinugawa$^{2}$}%
 \email{kinugawa@cc.nara-wu.ac.jp}
\affiliation{
$^{1}$Japan Science and Technology Corporation
\\and\\
$^{2}$Department of Chemistry, Faculty of Science, 
Nara Women's University,
 \\
Nara 630-8506, Japan}

\date{\today}

\begin{abstract}
We propose a new quantum dynamics method called 
the effective potential analytic continuation (EPAC)
to calculate the real time quantum correlation functions 
at finite temperature.
The method is based on the effective action formalism 
which includes the standard effective potential.
The basic notions of the EPAC are presented 
for a one-dimensional double well system
in comparison with 
the centroid molecular dynamics (CMD) 
and the exact real time quantum correlation function.
It is shown that both the EPAC and the CMD 
well reproduce the exact
short time behavior, while at longer time their results
deviate from the exact one. 
The CMD correlation function damps rapidly with time  
because of ensemble dephasing.    
The EPAC correlation function, however,  
can reproduce the long time oscillation 
inherent in 
the quantum double well systems.
It is also shown that
the EPAC correlation function
can be improved toward 
the exact correlation function
by means of 
the higher order derivative expansion of 
the effective action.
\end{abstract}
\maketitle
\section{INTRODUCTION}
\hspace*{\parindent}
The imaginary time path integral \cite{fh} provides 
the quantum statistical mechanical formalism 
of useful numerical methods 
for quantum many-body systems.
Static observables in such systems have successfully 
been calculated by the 
path integral Monte Carlo (PIMC) or 
path integral molecular dynamics (PIMD) 
calculation \cite{bt,ce}.
However, these numerical methods have difficulty 
in accessing the dynamical properties such as 
the real time quantum correlation functions,
because the analytic continuation from the imaginary time 
to the real time \cite{bm,agd,tb}
using a finite number of noisy imaginary time PIMC/PIMD data 
is so nontrivial as to be classified as 
an ill-posed problem.
To overcome such difficulty, 
the numerical analytic continuation scheme based on 
the maximum entropy method has been proposed to be applied
to various quantum dynamical problems \cite{si,gb,nah}.
As an alternative approach, one can directly evaluate 
the real time path integrals, although this
suffers from {\it the sign problem}; 
the error grows exponentially with time 
because of the rapid oscillation which 
originates from the factor $\exp(iS[q(t)]/\hbar)$ \cite{bt,tm,me}. 
In addition to these approaches, another numerical method 
has been proposed
to evaluate the eigenstates of quantum systems using the path integral 
and to construct the real time quantum correlation functions
\cite{hirata}.
\par
The effective potential is a device widely used 
for the approximate calculations of static 
properties of quantum many-body systems. 
There are several definitions of effective potentials: 
{\it effective classical potential} ~\cite{fh,fk,gt}, 
{\it standard effective potential} ~\cite{ep}, 
and so on \cite{st,sv,ok,jklmr}.
All of them are defined in the framework of path integral 
while the relations among them 
have been discussed so far \cite{fk,fukuda,hs,owy,ks}.
The usefulness of the effective potentials 
has been indicated by many applications to
simple quantum mechanical systems \cite{fk,gt},
condensed phase systems \cite{ctvv},
the quantum transition-state theory \cite{gil,vm}, 
and quantum field theories \cite{co}.
It is true that static properties of the systems 
are well described in terms of 
the effective potentials in a simple classical analogue.
For instance, the quantum-mechanical partition function
is expressed as a classical-like partition function 
including the {\it effective classical potential}.
Furthermore, both of the thermodynamic phases 
of quantum statistical systems and 
the vacuum structure of quantum field theories 
are determined by 
the minima of the {\it standard effective potentials}.
However, it has been believed that such 
classical use of the effective potentials cannot 
be directly applied to the dynamical problems
because the definitions of the effective potentials,
in most cases, do not justify the use of the classical equations 
of motion on the effective potential surface.
Therefore, for the calculation of time-dependent properties 
such as the real time quantum correlation functions,
more careful treatment of the effective potentials is required.
\par
On the other hand, Cao and Voth have recently proposed the 
{\it centroid molecular dynamics} (CMD) method, which is
an approximation to obtain real time quantum correlation functions
from molecular dynamics on the {\it effective classical potential}
surface \cite{cv}.
The validity of this method is ensured by the fact that 
the real time correlation function of the centroid variables is 
a good approximation to the exact canonical correlation function
in the linear response theory \cite{kubo,jv}.  
The CMD is a promising method suited for 
the numerical computation of the dynamics of 
many-body molecular systems
such as condensed matter and molecular clusters \cite{appli}.
In fact, the validity of the CMD approximation has been tested 
for low dimensional nonlinear systems;    
the real time correlation functions obtained from the CMD 
are more accurate at shorter time and at lower temperature,
whereas they are evidently worse at longer time \cite{jv,kb}.   
It has also been found that the time correlation function
evaluated by the CMD damps rapidly with time    
because of {\it ensemble dephasing}, 
which is well-known behavior in one-dimensional 
nonlinear classical systems \cite{jv}.
\par
In the present paper, we propose 
a new quantum dynamics method, called
the {\it effective potential analytic continuation} (EPAC) method,
to calculate the real time quantum correlation functions. 
The EPAC method is based on 
the effective action formalism \cite{ep,riv,swa,ps,kl}, where 
imaginary time quantum correlation functions are 
expressed in terms of the effective action.
This method is an approximation method
which includes the {\it standard effective potential}
defined as the leading order of the derivative expansion
of the effective action.
Once the {\it standard effective potential} is known, 
one can obtain the analytic form of 
the imaginary time quantum correlation function,
and then the analytic continuation from the imaginary time to
the real time can be readily performed.
The EPAC is expected to be a 
powerful method to calculate 
the real time quantum correlation functions
with less computational effort
than the other numerical methods, because
the {\it standard effective potential} 
can be easily calculated 
from the {\it effective classical potential} ~\cite{owy}.
\par
In the present work, we apply the EPAC method to 
a one-dimensional quantum double well system at finite temperature.
At first, the {\it standard effective potential} is 
obtained from the {\it effective classical potential} calculated 
by means of the PIMD technique.
And then the 
real time quantum correlation function is constructed 
by means of the analytic continuation procedure. 
The results are compared with 
the exact quantum statistical mechanical results 
to investigate the accuracy of the EPAC method.
We also compare the EPAC results with the CMD results,
to clarify the difference between
the approximate quantum dynamics based 
on the {\it standard effective potential} and 
on the {\it effective classical potential}.
Finally a possible improvement of the EPAC is discussed
on the basis of the derivative expansion. 
\par
In Sec. II, we summarize the definitions and fundamental properties 
of two types of the effective potentials.
In Sec. III, the CMD method is briefly surveyed.
We then newly introduce the EPAC method and 
present its numerical implementation.
In Sec. IV, the results of the numerical tests 
for the accuracy of the EPAC method are shown 
in comparison with 
the CMD and the exact results.
A possible improvement of the EPAC method is also discussed.
The conclusions are given in Sec. V.
\par
Throughout this work, 
we consider a quantum particle of mass $m$ in a one-dimensional 
potential $V(q)$ at temperature $T$.
\section{DEFINITIONS OF THE EFFECTIVE POTENTIALS}
\subsection{Effective classical potential}
\hspace*{\parindent}
At first, we summarize Feynman's definition  
of the effective classical potential \cite{fh}.
The quantum canonical partition function
for a one-dimensional system 
at temperature $T$ is expressed
in terms of the imaginary time path integral
\begin{eqnarray}
{\cal Z}_{\beta}&=&\int^{\infty}_{-\infty}dq
\int^{q(\beta\hbar)=q}_{q(0)=q}{\cal D}q
~\!\exp\left[~\!-~\!\frac{1}{\hbar}
~\!S_{E}[q]~\!\right]
, \label{1}
\end{eqnarray}
where $\beta=1/k_{B}T$ and $S_{E}[q]$ is the
Euclidean action functional
\begin{eqnarray}
S_{E}[q]=\int^{\beta\hbar}_{0}d\tau
\left[~\!\frac{1}{2}~\!m ~\!\dot{q}^2 + V(q)~\!\right]
. \label{2}
\end{eqnarray}
The imaginary time average value of $q(\tau)$ is defined as
\begin{eqnarray}
q_0=\frac{1}{\beta\hbar}\int^{\beta\hbar}_{0}d\tau ~\!q(\tau)
. \label{3}
\end{eqnarray}
This is the zero mode in 
the Fourier modes of $q(\tau)$, 
$q_n=(1/\beta\hbar)\int^{\beta\hbar}_{0}d\tau 
e^{i(2\pi n/\beta\hbar)\tau}~\!q(\tau)$,
and is referred to as the {\it path centroid}.
Inserting $1=\int^{\infty}_{-\infty}dq_{c}\delta(q_{0}-q_{c})$ 
to the integral of Eq. (\ref{1}), we obtain 
\begin{eqnarray}
{\cal Z}_{\beta}&=&
\int^{\infty}_{-\infty}dq_{c}\int^{\infty}_{-\infty}dq
\int^{q(\beta\hbar)=q}_{q(0)=q}{\cal D}q
~\delta(q_{0}-q_{c})
~\!e^{-~\!S_{E}/\hbar}
\nonumber\\
&=&\int^{\infty}_{-\infty}dq_{c}
~\!\rho_{\beta}^{c}(q_{c})
\nonumber\\
&=&\sqrt{\frac{m}{2\pi\beta\hbar^2}}
\int^{\infty}_{-\infty}dq_{c}
~\!e^{-\beta V_{\beta}^{c}(q_{c})}
.\label{6}
\end{eqnarray}
Here the centroid density
\begin{eqnarray}
\rho_{\beta}^{c}(q_{c})&=&
\int^{\infty}_{-\infty}dq
\int^{q(\beta\hbar)=q}_{q(0)=q}{\cal D}q
~\delta(q_{0}-q_{c})
~\!e^{-~\!S_{E}/\hbar}
,\label{7}
\end{eqnarray}
and the {\it effective classical potential}
\begin{eqnarray}
V_{\beta}^{c}(q_{c})&=&
-\frac{1}{\beta}\log\left[
\sqrt{\frac{2\pi\beta\hbar^2}{m}}~\!\rho_{\beta}^{c}(q_{c})
\right]
,\label{8}
\end{eqnarray}
have been introduced as the functions
of {\it position centroid variable} $q_{c}$.
This type of effective potential, Eq. (\ref{8}),
contains the effects of quantum fluctuations
of $q_{n\ne 0}$ modes  
and is also called 
{\it effective centroid potential} ~\cite{cv},
{\it constraint effective potential} ~\cite{owy},
or {\it Wilsonian effective potential} ~\cite{wk,wh}.
This effective potential can be numerically 
evaluated by using the PIMC or the PIMD techniques \cite{hs,owy,ks,cv}.  
\subsection{Standard effective potential}
\hspace*{\parindent}
In this subsection the effective action formalism 
is briefly reviewed and the standard effective potential 
is then introduced 
by means of the derivative expansion of the effective action.
All the contents shown below 
can be seen in Refs. \onlinecite{riv,swa,ps,kl}.
\par 
Consider an imaginary time quantum theory in the presence 
of an external source $J(\tau)$.
The quantum canonical partition function 
of a one-dimensional quantum system with $J(\tau)$
is expressed as
\begin{eqnarray}
{\cal Z}_{\beta}[J]&=&\int^{\infty}_{-\infty}dq
\int^{q(\beta\hbar)=q}_{q(0)=q}{\cal D}q
~\!\exp\left[~\!-~\!\frac{1}{\hbar}
~\!S_{E}[q]+\frac{1}{\hbar}
~\!\int^{\beta\hbar}_{0}d\tau J(\tau)q(\tau)
\right]
,\label{9}
\end{eqnarray}
while the generating functional is defined as 
\begin{eqnarray}
W_{\beta}[J]
&=& \hbar \log {\cal Z}_{\beta}[J] 
.\label{10}
\end{eqnarray}
The functional derivatives of $W_{\beta}[J]$ with respect to $J(\tau)$ 
produce the connected Green functions in the presence of $J(\tau)$
\begin{eqnarray}
\frac{\delta W_{\beta}[J]}{\delta J(\tau)}
&=&\langle \hat{q}(\tau)\rangle_{\beta}^{J} \equiv Q(\tau) 
,\label{11}\\
\hbar\frac{\delta^{2} W_{\beta}[J]}{\delta J(\tau)\delta J(s)}
&=&\langle T \hat{q}(\tau)\hat{q}(s)\rangle_{\beta}^{J}
-\langle \hat{q}(\tau)\rangle_{\beta}^{J}
~\!\langle \hat{q}(s)\rangle_{\beta}^{J}
,\label{12}
\end{eqnarray}
where $T$ represents the time-ordered product.
Then the imaginary time connected Green function 
$G_{\beta}^{\rm conn}(\tau-s)$ 
in thermal equilibrium is obtained when 
the external source $J$ vanishes
\begin{eqnarray}
G_{\beta}^{{\rm conn}}(\tau-s)&=&
\left.\hbar\frac{\delta^{2} W_{\beta}[J]}{\delta J(\tau)\delta J(s)}
\right|_{J=0}
=G_{\beta}(\tau-s)
-\langle \hat{q}(\tau)\rangle_{\beta}
~\!\langle \hat{q}(s)\rangle_{\beta}
,\label{13}
\end{eqnarray}
where $G_{\beta}(\tau-s)$
is the imaginary time Green function 
(the Matsubara Green function)
\begin{eqnarray}
G_{\beta}(\tau-s)
&=&\langle T \hat{q}(\tau)\hat{q}(s)\rangle_{\beta}
=\theta(\tau-s)~\langle \hat{q}(\tau)\hat{q}(s)\rangle_{\beta}
+\theta(s-\tau)~\langle \hat{q}(s)\hat{q}(\tau)\rangle_{\beta}
.\label{13a}
\end{eqnarray}
Note that the expectation value $\langle \hat{q}(\tau)\rangle_{\beta}$
is independent of the imaginary time in thermal equilibrium.
The {\it effective action} $\Gamma_{\beta}[Q]$ is defined 
by the Legendre transform of $W_{\beta}[J]$
\begin{eqnarray}
\Gamma_{\beta}[Q]&=&-W_{\beta}[J]
+\int^{\beta\hbar}_{0}d\tau J(\tau)Q(\tau)
.\label{14}
\end{eqnarray}
This satisfies the equation
\begin{eqnarray}
\frac{\delta\Gamma_{\beta}[Q]}{\delta Q(\tau)}&=&J(\tau)
,\label{15}
\end{eqnarray}
which becomes the quantum mechanical Euler-Lagrange equation 
in the $J\to 0$ limit, i.e., 
the principle of the least action.
The functional derivative of Eq. (\ref{15}) with respect to $J(s)$ 
leads to 
\begin{eqnarray}
\int^{\beta\hbar}_{0}d u ~
\frac{\delta^2 W_{\beta}[J]}{\delta J(s)\delta J(u)}~
\frac{\delta^2\Gamma_{\beta}[Q]}{\delta Q(u)\delta Q(\tau)}
&=&\delta(\tau-s)
.\label{16}
\end{eqnarray}
When the Fourier mode expansions with 
the Matsubara frequencies $\omega_{n}=2\pi n/\beta\hbar$
are defined, we get
\begin{eqnarray}
G_{\beta}^{\rm conn}(\tau-s)&=&\frac{1}{\beta\hbar}
\sum^{\infty}_{n=-\infty}e^{-i\omega_{n}(\tau-s)}
G^{\rm conn}_{\beta}(\omega_{n})
,\label{17}\\
\frac{\delta^2 \Gamma_{\beta}[Q]}{\delta Q(\tau)\delta Q(s)}
&=&\frac{1}{\beta\hbar}
\sum^{\infty}_{n=-\infty}e^{-i\omega_{n}(\tau-s)}
\frac{\delta^2 \Gamma_{\beta}}{\delta Q\delta Q}(\omega_n)
.\label{18}
\end{eqnarray}
Then a general relation is obtained from Eq. (\ref{16})
in the $J\to 0$ limit
\begin{eqnarray}
G_{\beta}^{\rm conn}(\omega_n)&=&\left(
\frac{1}{\hbar}\frac{\delta^2\Gamma_{\beta}}{\delta Q\delta Q}(\omega_n)
\right)^{-1}
.\label{19}
\end{eqnarray}
Thus the imaginary time (two-point) Green function is obtained
with the knowledge of the effective action $\Gamma_{\beta}[Q]$.
In a similar way, 
the imaginary time $n$-point Green function
can be obtained with the knowledge 
of $n$th functional derivative of the effective action.
Therefore, the effective action formalism reviewed here
provides a powerful scheme to calculate 
the correlation functions.
\par
In general, the effective action $\Gamma_{\beta}[Q]$ is nonlocal 
in the imaginary time $\tau$, so that 
it is formally possible to expand $\Gamma_{\beta}[Q]$ 
in a series of the terms 
involving imaginary time derivatives of $q(\tau)$
(the {\it derivative expansion})
\begin{eqnarray}
\Gamma_{\beta}[Q]&=&\int^{\beta\hbar}_{0}d\tau\left[
V_{\beta}(Q)+\frac{1}{2}Z_{\beta}(Q)\dot{Q}^{2}+{\it O}(\partial^{4})
\right]
,\label{20}
\end{eqnarray}
where $V_{\beta}(Q)$ is called the {\it standard effective potential},
the leading order of the derivative expansion of $\Gamma_{\beta}[Q]$.
\par
A significant feature of $V_{\beta}(Q)$ 
is its {\it convexity}, which can be easily shown below.
When we set $Q(\tau)\to Q$: constant,
the effective action is written as  
$\Gamma_{\beta}[Q]=\beta\hbar V_{\beta}(Q)$.
In a similar way, when we set $J(\tau)\to J$: constant, 
the generating functional $W_{\beta}[J]$  
can be written in terms of the density $w(J)$: 
$W_{\beta}[J]=\beta\hbar w(J)$.
Then Eqs. (\ref{15}) and (\ref{16}) become
\begin{eqnarray}
\frac{\partial  V_{\beta}(Q)}{\partial Q}&=&J
,\label{21a}\\
\frac{1}{\beta\hbar}\frac{\partial^2 w_{\beta}(J)}{\partial J^2}~
\frac{\partial^2 V_{\beta}(Q)}{\partial Q^2}&=&1
.\label{21b}
\end{eqnarray}
Considering the inequality 
\begin{eqnarray}
\frac{\partial^2 w_{\beta}(J)}{\partial J^2}&=&
\frac{(\beta\hbar)^{2}}{\hbar}\langle~\!
\left(~\!\hat{q}_{0}-\langle \hat{q}_{0}~\!\rangle_{\beta}
\right)^{2}\rangle_{\beta}\geq 0
,\label{22}
\end{eqnarray}
one obtains 
\begin{eqnarray}
\frac{\partial^2 V_{\beta}(Q)}{\partial Q^2}\geq 0
.\label{23}
\end{eqnarray}
Therefore, the standard effective potential $V_{\beta}(Q)$ 
is always convex, while
the effective classical potential $V_{\beta}^{c}(q_{c})$ 
is not necessarily convex.
\subsection{Relationship between $V^{c}_{\beta}(q_c)$ and $V_{\beta}(Q)$ }
\hspace*{\parindent}
In this subsection we mention the relationship between 
the effective classical potential $V^{c}_{\beta}(q_c)$ and 
the standard effective potential $V_{\beta}(Q)$ \cite{fukuda,hs,owy,ks}.
Replacing $\delta(q_{0}-q_{c})$ in Eq. (\ref{7}) by its integral
representation
\begin{eqnarray}
\delta(q_{0}-q_{c})&=&\frac{\beta}{2\pi i}\int^{c+i\infty}_{c-i\infty}
d J ~\!e^{\beta J (q_{0}-q_{c})}
,\label{24}
\end{eqnarray}
one obtains
\begin{eqnarray}
e^{-\beta V_{\beta}^{c}(q_{c})}&=&
\sqrt{\frac{2\pi\beta\hbar^2}{m}}\frac{\beta}{2\pi i}
\int^{c+i\infty}_{c-i\infty} d J
\int^{\infty}_{-\infty}dq
\int^{q(\beta\hbar)=q}_{q(0)=q}{\cal D}q
~e^{\beta J (q_{0}-q_{c})}e^{-S_{E}/\hbar}
\nonumber\\
&=&\sqrt{\frac{2\pi\beta\hbar^2}{m}}\frac{\beta}{2\pi i}
\int^{c+i\infty}_{c-i\infty} d J
~e^{-\beta ~[-w_{\beta}(J)+J q_{c}]}
.\label{26}
\end{eqnarray}
When the low temperature limit $\beta\to \infty$ is taken, 
the integral $\int dJ$ can be evaluated  
by the saddle-point method
\begin{eqnarray}
e^{-\beta V_{\beta}^{c}(q_{c})}&=& C 
~e^{-\beta ~\!\left.[-w_{\beta}(J)+J q_{c}]~\!
\right |_{q_{c}=\frac{\partial w_{\beta}}{\partial J}}}
\nonumber\\
&=& C ~e^{-\beta V_{\beta}(q_c)}
,\label{28}
\end{eqnarray}
where $C$ is a constant. 
Therefore these two effective potentials are equal
\begin{eqnarray}
\lim_{\beta\to \infty} V_{\beta}^{c}(q_{c})&=& 
\lim_{\beta\to \infty} V_{\beta}(q_{c})
,\label{29}
\end{eqnarray}
except for an additive constant term.
Namely, both the effective potentials coincide 
with each other at the zero temperature.
This relation also ensures the convexity of 
the effective classical potential $V_{\beta}^{c}(q_{c})$
in the low temperature limit.
\section{REAL TIME CORRELATION FUNCTIONS}
\hspace*{\parindent}
Now we proceed to the calculations of    
the real time quantum correlation functions
starting from the effective potentials.
First we briefly review the CMD method \cite{cv}
which is an approximation using 
the effective classical potential $V_{\beta}^{c}(q_{c})$.
Then we newly introduce our EPAC method,
which is a novel approximation using 
the standard effective potential $V_{\beta}(Q)$.
In this section, we concentrate on 
the two-point position correlation function
\begin{eqnarray}
C_{\beta}(t)&=&\langle \hat{q}(t)\hat{q}(0)\rangle_{\beta},
\label{30}
\end{eqnarray}
to clarify the essence of the approximations.
\subsection{Centroid molecular dynamics method}
\hspace*{\parindent}
In the CMD, a {\it real time} classical equation of motion 
for the {\it position centroid variable} $q_{c}(t)$
on the effective classical potential $V_{\beta}^{c}(q_{c})$
is introduced
\begin{eqnarray}
m~\!\ddot{q}_{c}(t)&=&F_{\beta}^{c}(q_{c})=
-\frac{\partial V_{\beta}^{c}(q_{c})}{\partial q_{c}}
.\label{31}
\end{eqnarray}
The centroid force $F_{\beta}^{c}(q_{c})$ is evaluated by 
the Ehrenfest relation for $q_{c}$ \cite{owy}
\begin{eqnarray}
F_{\beta}^{c}(q_{c})&=&-\frac{1}{\rho_{\beta}^{c}}
\int^{\infty}_{-\infty}dq
\int^{q(\beta\hbar)=q}_{q(0)=q}{\cal D}q
~\delta(q_{0}-q_{c})
\left[\frac{1}{\beta\hbar}\int^{\beta\hbar}_{0}d s 
\frac{\partial V(q)}{\partial q(s)}\right]
~\!e^{-~\!S_{E}/\hbar}
\nonumber\\
&=&-\left<\frac{1}{\beta\hbar}\int^{\beta\hbar}_{0}d s 
\frac{\partial \hat{V}(q)}{\partial q(s)}\right>_{\beta}^{c}
,\label{33}
\end{eqnarray}
where $\langle\cdot\cdot\cdot\rangle^{c}_{\beta}$ denotes the 
quantum mechanical average with a constraint
$q_{0}=q_{c}$. 
Using the centroid trajectory $q_{c}(t)$ generated  
from Eq. (\ref{31}), one can construct the  
{\it centroid correlation function} in a classical fashion
\begin{eqnarray}
C_{\beta}^{c}(t)&=&\langle q_{c}(t)q_{c}(0)\rangle_{\rho_{\beta}^{c}}
=\frac{1}{{\cal Z}_{\beta}}
\int^{\infty}_{-\infty}dq_{c}~q_{c}(t)q_{c}(0)~\rho_{\beta}^{c}(q_{c})
\nonumber\\
&=&\frac{1}{{\cal Z}_{\beta}}
\int^{\infty}_{-\infty}\int^{\infty}_{-\infty}
\frac{dq_{c}dp_{c}}{2\pi\hbar}~q_{c}(t)q_{c}(0)~
e^{-\beta [p_{c}^{2}/2m+V_{\beta}^{c}(q_{c})]}
.\label{35}
\end{eqnarray}
Here we introduced the {\it momentum centroid variable} $p_{c}(t)$.
Cao and Voth proposed the CMD approximation \cite{cv}, 
in which an approximate relation holds
between two correlation functions
\begin{eqnarray}
C_{\beta}^{c}(t)&\simeq&C_{\beta}^{\rm CAN}(t)
,\label{36}
\end{eqnarray}
where $C_{\beta}^{\rm CAN}(t)$ is the 
canonical correlation function 
appearing in the linear response theory \cite{kubo}
\begin{eqnarray}
C_{\beta}^{\rm CAN}(t)&=&\frac{1}{\beta}\int_{0}^{\beta}d\lambda
~\langle \hat{q}(t-i\hbar\lambda)\hat{q}(0)\rangle_{\beta}
.\label{37}
\end{eqnarray}
In fact, the canonical correlation function is related to 
the real time quantum correlation function in the Fourier space
\begin{eqnarray}
C_{\beta}(\omega)&=&E(\omega)~
C_{\beta}^{\rm CAN}(\omega)
,\label{38a}
\end{eqnarray}
where we introduced a function
\begin{eqnarray}
E(\omega)&=&\frac{\beta\hbar\omega}{2}\left(
\coth\frac{\beta\hbar\omega}{2}+1\right)
.\label{38b}
\end{eqnarray}
Therefore, the CMD method enables us to evaluate 
quantum dynamics 
from the correlation function $C_{\beta}^{c}(t)$ calculated  
in a classical manner. 
The relation Eq. (\ref{36}) is exact for a harmonic oscillator.
For general potentials, $C_{\beta}^{c}$ is exactly equal to
$C_{\beta}^{\rm CAN}$ 
in the classical limit or
for linear operators at $t=0$,
as shown in Appendix A.
\par
For general correlation functions of 
nonlinear operators, e.g.
$\langle \hat{q}^n(t)\hat{q}^n(0)\rangle_{\beta}$,   
the relation Eq. (\ref{36}) is not valid even at $t=0$.
Reichman et al. have investigated this problem 
to propose the use of the higher-order Kubo transforms \cite{rrjv}. 
\subsection{Effective potential analytic continuation method}
\hspace*{\parindent}
In this subsection we newly present our quantum dynamics method.
We begin with the review of the standard procedure 
of the analytic continuation from the imaginary time 
to the real time \cite{kl,bell}.
Let us start from the Fourier coefficient of 
the Matsubara Green function [Eq. (\ref{13a})]
\begin{eqnarray}
G_{\beta}(\omega_{n})
&=&\int^{\beta\hbar}_{0} 
d\tau e^{i\omega_{n}\tau}G_{\beta}(\tau)
.\label{39b}
\end{eqnarray}
Next the real time quantities,
the retarded and advanced Green functions, are introduced
\begin{eqnarray}
G_{\beta}^{R}(t)
&=&~\theta(t)~\langle ~[\hat{q}(t),\hat{q}(0)]~
\rangle_{\beta}
,\label{39c}\\
G_{\beta}^{A}(t)
&=&\!\!\theta(-t)~\langle ~[\hat{q}(0),\hat{q}(t)]~
\rangle_{\beta}
,\label{39d}
\end{eqnarray}
where $[~,~]$ denotes the commutator. 
The Fourier coefficients of 
these Green functions are 
$G_{\beta}^{R}(\omega)=\int^{\infty}_{-\infty} 
dt~ e^{i\omega t}G_{\beta}^{R}(t)$ and 
$G_{\beta}^{A}(\omega)=\int^{\infty}_{-\infty} 
dt~ e^{i\omega t}G_{\beta}^{A}(t)$, respectively.
One also introduces a couple of correlation functions,
$C_{\beta}^{>}(t)=C_{\beta}(t)=
\langle \hat{q}(t)\hat{q}(0)\rangle_{\beta}$, 
$C_{\beta}^{<}(t)=
\langle \hat{q}(0)\hat{q}(t)\rangle_{\beta}$, 
and their Fourier coefficients,
$C_{\beta}^{>}(\omega)=\int^{\infty}_{-\infty} 
dt~ e^{i\omega t}C_{\beta}^{>}(t)$, 
$C_{\beta}^{<}(\omega)=\int^{\infty}_{-\infty} 
dt~ e^{i\omega t}C_{\beta}^{<}(t)$.
When one defines the spectral function 
\begin{eqnarray}
\rho_{\beta}(\omega)
&=&\int^{\infty}_{-\infty} 
dt~ e^{i\omega t}\langle ~[\hat{q}(t),\hat{q}(0)]~
\rangle_{\beta}
\label{39e}\\
&=&~C_{\beta}^{>}(\omega)-C_{\beta}^{<}(\omega)
\label{39f}\\
&=&~(1-e^{-\beta\hbar\omega})C_{\beta}^{>}(\omega)
,\label{39f2}
\end{eqnarray}
then the quantities $G_{\beta}(\omega_{n})$, $G_{\beta}^{R}(\omega)$,
and $G_{\beta}^{A}(\omega)$ are expressed in terms of it
\begin{eqnarray}
G_{\beta}(\omega_{n})
&=&~-\int^{\infty}_{-\infty} 
\frac{d\omega '}{2\pi}~ \frac{1}{i\omega_{n}-\omega '}
~\rho_{\beta}(\omega ')
,\label{39g}\\
G_{\beta}^{R}(\omega)
&=&~i\int^{\infty}_{-\infty} 
\frac{d\omega '}{2\pi}~ \frac{1}{\omega-\omega '+i\epsilon}
~\rho_{\beta}(\omega ')
,\label{39h}\\
G_{\beta}^{A}(\omega)
&=&~i\int^{\infty}_{-\infty} 
\frac{d\omega '}{2\pi}~ \frac{1}{\omega-\omega '-i\epsilon}
~\rho_{\beta}(\omega ')
,\label{39i}
\end{eqnarray}
where $\epsilon$ is a positive infinitesimal.  
Therefore one can obtain the real time quantities 
$G_{\beta}^{R}(\omega)$ and $G_{\beta}^{A}(\omega)$
from the imaginary time quantity $G_{\beta}(\omega_{n})$
by means of the {\it analytic continuation}:
\begin{eqnarray}
G_{\beta}^{R}(\omega)
=-i \left.G_{\beta}(\omega_{n})\right|_{i\omega_{n}=\omega+i\epsilon},
~~~~G_{\beta}^{A}(\omega)
=-i \left.G_{\beta}(\omega_{n})\right|_{i\omega_{n}=\omega-i\epsilon} 
.\label{39j}
\end{eqnarray}
Since the spectral function $\rho_{\beta}(\omega)$ is also 
expressed as
\begin{eqnarray}
\rho_{\beta}(\omega)
&=&~G_{\beta}^{R}(\omega)-G_{\beta}^{A}(\omega)
,\label{39j2}
\end{eqnarray}
and 
the real time quantum correlation function 
$C_{\beta}(t)=\langle \hat{q}(t)\hat{q}(0)\rangle_{\beta}$
is expressed as 
\begin{eqnarray}
C_{\beta}(t)=
\int^{\infty}_{-\infty} 
\frac{d\omega}{2\pi}~ e^{-i\omega t}C_{\beta}^{>}(\omega)=
\int^{\infty}_{-\infty} 
\frac{d\omega}{2\pi}~ e^{-i\omega t}
\left[ 1+\frac{1}{e^{\beta\hbar\omega}-1}\right]
\rho_{\beta}(\omega)
,\label{39k}
\end{eqnarray}
one can obtain the real time quantum dynamics 
with the knowledge of the imaginary time quantity 
via the analytic continuation procedure.
\par
Now we present the EPAC method,
with which one can readily perform 
the analytic continuation procedure 
shown above, utilizing  
the effective action formalism \cite{riv,swa,ps,kl}. 
As seen in Eq. (\ref{19}),  
the imaginary time two-point connected  
Green function is expressed in terms of 
the effective action $\Gamma_{\beta}[Q]$.
Now we employ the local potential approximation
\begin{eqnarray}
\Gamma_{\beta}[Q]&=&\int^{\beta\hbar}_{0}d\tau\left[
V_{\beta}(Q)+\frac{1}{2}~\!m~\!\dot{Q}^{2}
\right]
,\label{39}
\end{eqnarray}
where any derivative terms are dropped except for 
the kinetic term.
From Eq. (\ref{39}), the second functional derivative 
of $\Gamma_{\beta}[Q]$
becomes the second derivative of 
the standard effective potential $V_{\beta}(Q)$
\begin{eqnarray}
\frac{\delta^2 \Gamma_{\beta}}{\delta Q(\tau)\delta Q(s)}
&=&\left(-m\frac{d^2}{d\tau^2}
+\frac{\partial^2 V_{\beta}}{\partial Q^2}\right)\delta(\tau-s)
,\label{40}\\
\frac{\delta^2 \Gamma_{\beta}}{\delta Q\delta Q}(\omega_n)
&=&m\omega_{n}^{2}+\frac{\partial^2 V_{\beta}}{\partial Q^2}
.\label{41}
\end{eqnarray}
Note that $\partial^2 V_{\beta}/\partial Q^2$ must be evaluated 
at $Q=Q_{\rm min}$ which gives 
the minimum of $V_{\beta}(Q)$. 
This is because Eq. (\ref{19}) holds
in the $J\to 0$ limit and the value of $Q$ is fixed 
at $Q=Q_{\rm min}$ in Eq. (\ref{21a}).
It should be also noted that $Q_{\rm min}$ is independent of
the imaginary time in thermal equilibrium.
When we define the effective frequency
\begin{eqnarray}
\omega_{\beta}&=&\sqrt{\frac{1}{m}\left.\frac{\partial^2 V_{\beta}}
{\partial Q^2}\right|}_{Q=Q_{\rm min}}
,\label{42}
\end{eqnarray}
from Eqs. (\ref{13}), (\ref{19}), (\ref{41}), and (\ref{42}),
the Fourier transformed Matsubara Green function  
$G_{\beta}(\omega_{n})$ is written as 
\begin{eqnarray}
G_{\beta}(\omega_{n})&=& 
G_{\beta}^{\rm conn}(\omega_{n})
=\frac{\hbar}{m\omega^{2}_{n}+m\omega^{2}_{\beta}}
.\label{43}
\end{eqnarray}
Here we omitted the Fourier coefficient of the 
constant term $Q^{2}_{\rm min}$ because it 
has nothing to do with the procedure described below.
\par
Following Eq. (\ref{39j}),
we obtain the real time quantities 
from the imaginary time quantity $G_{\beta}(\omega_{n})$
by the analytic continuation:
\begin{eqnarray}
G_{\beta}^{R}(\omega)
&=&
i\frac{\hbar}{2m\omega_{\beta}}
\left[\frac{1}{\omega-\omega_{\beta}+i\epsilon}
-\frac{1}{\omega+\omega_{\beta}+i\epsilon}
\right]
,\label{43a}\\
G_{\beta}^{A}(\omega)
&=&
i\frac{\hbar}{2m\omega_{\beta}}
\left[\frac{1}{\omega-\omega_{\beta}-i\epsilon}
-\frac{1}{\omega+\omega_{\beta}-i\epsilon}
\right]
.\label{43b}
\end{eqnarray}
Then, using Eq. (\ref{39j2}), the spectral function has the form 
\begin{eqnarray}
\rho_{\beta}(\omega)
&=&\frac{\pi\hbar}{m\omega_{\beta}}
\left[~\delta(\omega-\omega_{\beta})
-\delta(\omega+\omega_{\beta})~\right]
.\label{43c}
\end{eqnarray}
Using Eq. (\ref{39k}), we obtain 
an approximate real time quantum correlation 
function $C_{\beta}^{\rm AC}(t)$
with the constant term $Q^{2}_{\rm min}$
\begin{eqnarray}
C_{\beta}(t)\simeq C^{\rm AC}_{\beta}(t)
&=&\left(\frac{\hbar}{2m\omega_{\beta}}
\coth\frac{\beta\hbar\omega_{\beta}}{2}\right)\cos\omega_{\beta}t
-i\left(~\!\frac{\hbar}{2m\omega_{\beta}}\right)\sin\omega_{\beta}t
+Q_{\rm min}^{2}
.\label{46}
\end{eqnarray}
We call this novel procedure 
the {\it effective potential analytic continuation} 
(EPAC) method, an approximation based on 
the effective action formalism and the derivative expansion. 
The EPAC correlation function $C^{\rm AC}_{\beta}(t)$ is similar   
to the exact quantum correlation function of a harmonic oscillator.
However, note that the frequency $\omega_{\beta}$ is 
the result coming from the curvature of the 
standard effective potential $V_{\beta}(Q)$ at $Q=Q_{\rm min}$.
It should be further
noted that the EPAC differs from the 
{\it analytically continued effective harmonic theory}~\cite{cv} 
which is based 
on the centroid variables.
In this effective harmonic theory,
the optimized frequency $\bar{\omega}_{c}(q_{c})$ 
appearing in the time correlation functions
comes from the curvature of the 
effective classical potential $V_{\beta}^{c}(q_{c})$.
Therefore, such effective frequency 
$\bar{\omega}_{c}(q_{c})$ can be imaginary 
in some cases because $V_{\beta}^{c}(q_{c})$ is not 
always convex \cite{eht}.
\par
\begin{figure}[htb]
{
\begin{center}
\setlength{\unitlength}{0.4mm}
\begin{picture}(170,230)
\put(10,190){\framebox(65,40){%
 \shortstack{imaginary time \\ \\
             path integral \\ \\
${\cal Z}_{\beta}=\int{\cal D}q e^{- S_{E}/\hbar}$
}}}
\put(15,120){\framebox(45,45){%
 \shortstack{effective \\ \\ 
             classical \\ \\
             potential \\ \\
$V_{\beta}^{c}(q_{c})$
}}}
\put(103,120){\framebox(45,45){%
 \shortstack{standard  \\ \\ 
             effective \\ \\
             potential \\ \\
$V_{\beta}(Q)$
}}}
\put(10,65){\framebox(65,30){%
 \shortstack{CMD \\ \\ 
$C_{\beta}^{c}(t)\simeq C_{\beta}^{\rm CAN}(t)$
}}}
\put(100,65){\framebox(65,30){%
 \shortstack{EPAC \\ \\ 
$C_{\beta}^{\rm AC}(t)\simeq C_{\beta}(t)$
}}}
\put(50,0){\framebox(80,40){%
 \shortstack{real time quantum\\ \\
             correlation function\\ \\ \\ 
$C_{\beta}(t)=\langle \hat{q}(t)\hat{q}(0)\rangle_{\beta}$}}}
\put(35,190){\thicklines\vector(0,-1){24}}
\put(35,120){\thicklines\vector(0,-1){24}}
\put(120,120){\thicklines\vector(0,-1){24}}
\put(35,65){\thicklines\line(0,-1){40}}
\put(145,65){\thicklines\line(0,-1){40}}
\put(35,25){\thicklines\vector(1,0){14}}
\put(145,25){\thicklines\vector(-1,0){14}}
\put(60,145){\thicklines\vector(1,0){42}}
\put(40,175){\shortstack{PIMC/PIMD}}
\put(40,109){\shortstack{classical}}
\put(40,101){\shortstack{eq. of motion}}
\put(123,109){\shortstack{analytic}}
\put(123,101){\shortstack{continuation}}
\put(40,50){\shortstack{{\normalsize $C_{\beta}(\omega)=
E(\omega) ~\!C_{\beta}^{\rm CAN}(\omega)$}}}
\put(64,150){\shortstack{Legendre}}
\put(63,137){\shortstack{transform}}
\end{picture}
\end{center}
}
\vspace{0mm}
\caption{
Schematic diagram of the calculations of 
the real time quantum correlation function from
the effective potentials: 
The Cao-Voth CMD method and 
the presently proposed EPAC method.
}
 \label{fig:schemes}
\end{figure}
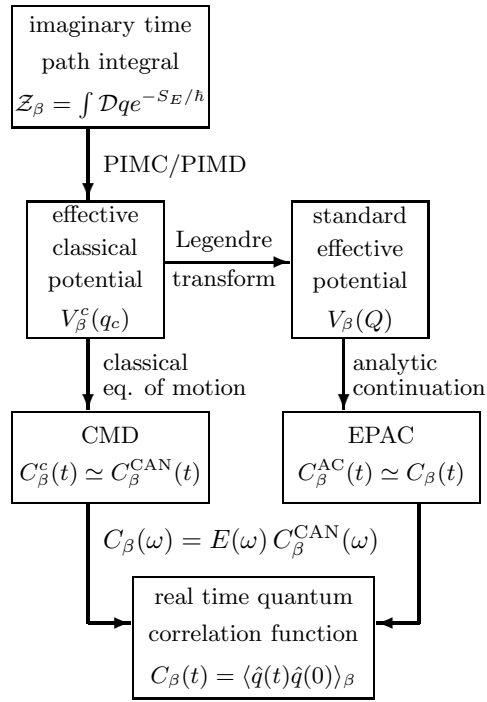
\par
To perform the EPAC procedure, one needs to 
calculate the standard effective potential $V_{\beta}(Q)$.
Once the effective classical potential $V_{\beta}^{c}(q_{c})$ 
is obtained, it is straightforward 
to calculate the standard effective potential as shown below \cite{owy}.
\par
From Eqs. (\ref{6}) and (\ref{10}), we have 
\begin{eqnarray}
e^{~\!\beta w_{\beta}(J)}&=&\sqrt{\frac{m}{2\pi\beta\hbar^2}}
\int^{\infty}_{-\infty}d q_{c}~\! 
e^{\beta~\![~\!J q_{c}-V_{\beta}^{c}(q_{c})~\!]},
\label{53}
\end{eqnarray}
with a constant source $J$. 
Therefore, we obtain $w_{\beta}(J)$ from $V_{\beta}^{c}(q_{c})$,
\begin{eqnarray}
w_{\beta}(J)&=&\frac{1}{\beta}\log\left[
\int^{\infty}_{-\infty}d q_{c}~\!e^{\beta 
~\![~\!J q_{c}-V_{\beta}^{c}(q_{c})~\!]}\right]+C,
\label{54}
\end{eqnarray}
where $C$  is a constant. 
As in Eq. (\ref{14}), 
$V_{\beta}(Q)$ is obtained through the Legendre transform 
\begin{eqnarray}
V_{\beta}(Q)&=&\sup_{J}~\!
\{~\!JQ-w_{\beta}(J)~\!\}
.\label{55}
\end{eqnarray}
Here we find the maximum value of $JQ-w_{\beta}(J)$ with varying 
$J$ for each given value of $Q$.
This definition is equivalent to Eq. (\ref{14}) 
though this is more suited for numerical evaluation.
\par
In Fig. \ref{fig:schemes} we show 
the flow chart of the calculations of   
the real time quantum correlation function 
by the CMD method and by the EPAC method.
\par
Finally we comment on the other contributed works 
of quantum dynamics methods based on  
the effective potentials. 
One of them is 
the direct use of the equation of motion Eq. (\ref{15}) 
in the $J\to 0$ limit \cite{ct}.  
Although this includes difficulty 
that Eq. (\ref{15}) contains (infinitely) 
many time derivatives, more detailed analyses 
have been performed \cite{cjlpt}.
On the other hand, 
it has been known that, 
in the zero temperature limit,
the CMD equation of motion [Eq. (\ref{31})]
can be obtained from Eq. (\ref{15}) in the $J\to 0$ limit,
employing the leading order derivative expansion 
and the analytic continuation. 
The meaning of the CMD equation 
in the zero temperature limit
has been discussed 
in connection with the minimum energy wave packet \cite{ra,vs},
while other related topics 
have also been discussed in Ref. \onlinecite{jk}.
\par
\section{NUMERICAL TESTS}
\hspace*{\parindent}
In this section we test the accuracy of the EPAC
method in a one-dimensional system 
by calculating the real time quantum correlation function 
$C_{\beta}^{\rm AC}(t)$ [Eq. (\ref{46})].
We compare $C_{\beta}^{\rm AC}(t)$ with 
the exact real time quantum correlation function
\begin{eqnarray}
C_{\beta}(t)=\frac{1}{{\cal Z}_{\beta}}\sum_{n}\sum_{m}
e^{-\beta E_{n}}e^{-i(E_{m}-E_{n})t/\hbar}
\left|\langle m|\hat{q}|n\rangle\right|^{2}
.\label{47}
\end{eqnarray}
For comparison, we also show the CMD results
of the centroid correlation function $C_{\beta}^{c}(t)$ 
[Eq. (\ref{35})]. 
In this case, $C_{\beta}^{c}(t)$ is compared with 
the exact canonical correlation function
\begin{eqnarray}
C_{\beta}^{\rm CAN}(t)=\frac{1}{{\cal Z}_{\beta}}\sum_{n}\sum_{m}
e^{-\beta E_{n}}\left[
\frac{1-e^{-\beta (E_{m}-E_{n})}}{\beta(E_{m}-E_{n})}
\right]e^{-i(E_{m}-E_{n})t/\hbar}
\left|\langle m|\hat{q}|n\rangle\right|^{2}
.\label{48}
\end{eqnarray}
The exact results of 
$C_{\beta}(t)$ and $C_{\beta}^{\rm CAN}(t)$ are obtained
from the eigenstates and the eigenvalues   
by solving the Schr\"odinger equation numerically.
Throughout the numerical evaluation, we employ natural units, 
$m=\hbar=k_{B}=1$.
\par
We consider a quantum particle moving on a
double well potential 
\begin{eqnarray}
V(q)=-\frac{1}{2}~\!q^{2}+\frac{1}{10}~\!q^{4}
.\label{49}
\end{eqnarray}
All the calculation have been performed at $\beta=1$ and $10$.
This double well potential model 
has already been analyzed in detail by use of the CMD 
and the other related methods \cite{jv}.
The parameters chosen here correspond to, for example, 
a proton potential where 
the potential height $\Delta V$ is 
about 0.5 kcal/mol
and the length between the potential minima $\Delta q$
is about 1 ${\rm \AA}$;
this is a relatively shallow double well potential model. 
The temperatures in the present study
are $T=1.6 \Delta V \simeq 400 [{\rm K}]$
for $\beta=1$ and     
$T=0.16 \Delta V \simeq 40 [{\rm K}]$
for $\beta=10$.
\par
In the following subsections,
first we show the CMD results 
with the exact canonical correlation function.
Next the EPAC results are compared  
with the exact quantum correlation function. 
We also present the Fourier transformed
correlation functions
to indicate the difference 
between the CMD method and the EPAC method clearly.
\subsection{Centroid molecular dynamics method}
\begin{figure}
\includegraphics[width=80mm,height=80mm]{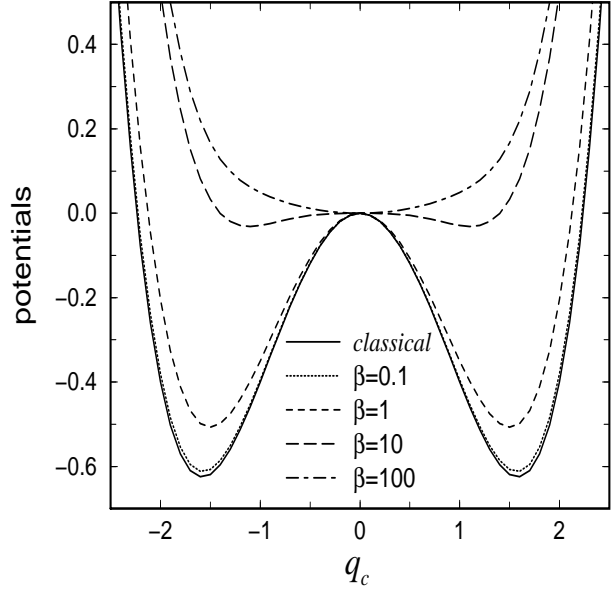}
\vspace{0mm}
\caption{
$\beta$-dependence of the effective classical potential 
$V_{\beta}^{c}(q_{c})$. We set $V_{\beta}^{c}(0)=0$.
}
\label{fig:ecp}
\end{figure}

\hspace*{\parindent}
For the CMD calculation, at first
we need to evaluate the 
centroid force $F_{\beta}^{c}(q_{c})$ [Eq. (\ref{33})]
before solving Eq. (\ref{31}) numerically.
For this purpose, we employed the PIMD technique.
The basis of the PIMD technique is 
the discretized representation of 
the quantum canonical partition function [Eq. (\ref{1})]
\begin{eqnarray}
{\cal Z}_{\beta}&=&\lim_{P\to \infty}
\left(\frac{mP}{2\pi\beta\hbar^2}\right)^{P/2}
\int\!\cdot\cdot\cdot\!\int
\prod _{j=1}^{P}dq_{j}~ e^{-\beta\Phi_{P}({\bf q})}
, \label{50}
\end{eqnarray}
where $P$ is the Trotter number and 
$\Phi_{P}({\bf q})$ is the potential of quasi-particles
${\bf q}=\{q_{1}, .. ,q_{P}\}$,
\begin{eqnarray}
\Phi_{P}({\bf q})&=&\sum^{P}_{j=1}
\left[\frac{mP}{2\beta^2\hbar^2}(q_{j}-q_{j+1})^2 
+ \frac{1}{P}V(q_{j})\right]
, \label{51}
\end{eqnarray}
with the periodic boundary condition $q_{P+1}=q_{1}$.
By performing the constant-temperature molecular dynamics (MD) 
for the quasi-particles,  
static properties corresponding to 
the quantum canonical partition function ${\cal Z}_{\beta}$ 
can be obtained as the averages over 
such MD-based configurations \cite{bt}. 
In this framework, the centroid force
\begin{eqnarray}
F_{\beta}^{c}(q_{c})&=&-\lim_{P\to \infty}
\frac{\int\!\cdot\cdot\cdot\!\int
\prod _{j=1}^{P}dq_{j}~ 
\delta\left(\frac{1}{P}\sum_{j=1}^{P}q_{j}-~q_{c}\right)
\left[\frac{1}{P}\sum_{j=1}^{P}
\frac{\partial V(q_{j})}{\partial q_{j}}\right]
e^{-\beta\Phi_{P}(\bf q)}}{
\int\!\cdot\cdot\cdot\!\int
\prod _{j=1}^{P}dq_{j}~ 
\delta\left(\frac{1}{P}\sum_{j=1}^{P}q_{j}-~q_{c}\right)
e^{-\beta\Phi_{P}(\bf q)}}
,\label{52}
\end{eqnarray}
was computed at each fixed centroid position $q_{c}$ 
by use of 
the normal mode PIMD algorithm \cite{cv,tmkp}
including the Nose-Hoover chain (NHC) thermostats \cite{mkt},
which ensure the generation of the static canonical distribution.
We used $10^7$ configurations 
to evaluate $F_{\beta}^{c}(q_{c})$ for 
each of 51 grid points of $q_{c}$,
equally spaced between $[-2.5, 2.5]$.  
Then we fitted the centroid force $F_{\beta}^{c}(q_{c})$
to the 25th polynomial function.
After this fitting, 
the effective classical potential $V_{\beta}^{c}(q_{c})$ 
was obtained from the integration of $F_{\beta}^{c}(q_{c})$. 
Figure \ref{fig:ecp} shows 
the classical (bare) double well potential $V(q)$ and  
the effective classical potential
$V_{\beta}^{c}(q_{c})$ at $\beta=0.1, 1, 10$ and $100$. 
As shown in Sec. II [Eqs. (\ref{23}) and (\ref{29})],
$V_{\beta}^{c}(q_{c})$ is not convex at higher temperature, 
though it becomes convex as the temperature lowers.
\par
\par
\par
The second step is to solve 
the CMD equation of motion [Eq. (\ref{31})] numerically. 
The centroid trajectories needed to compute 
the centroid correlation function $C_{\beta}^{c}(t)$ were 
calculated by means of the dual sampling procedure \cite{cv,jv,kb}.
In this procedure, at first the sampling of the initial position 
and momentum centroids was performed using  
the CMD with the centroid-attached NHC thermostat.
And then the centroid dynamics Eq. (\ref{31})
was computed from
the initial distribution in a {\it microcanonical} manner
to obtain the centroid trajectories.
The centroid correlation function was evaluated 
from such centroid trajectories spanning $10^{6}$ steps.
This is the correct evaluation of $C_{\beta}^{c}(t)$ following
its definition Eq. (\ref{35}).
\par
\begin{figure}
\includegraphics[width=80mm,height=80mm]{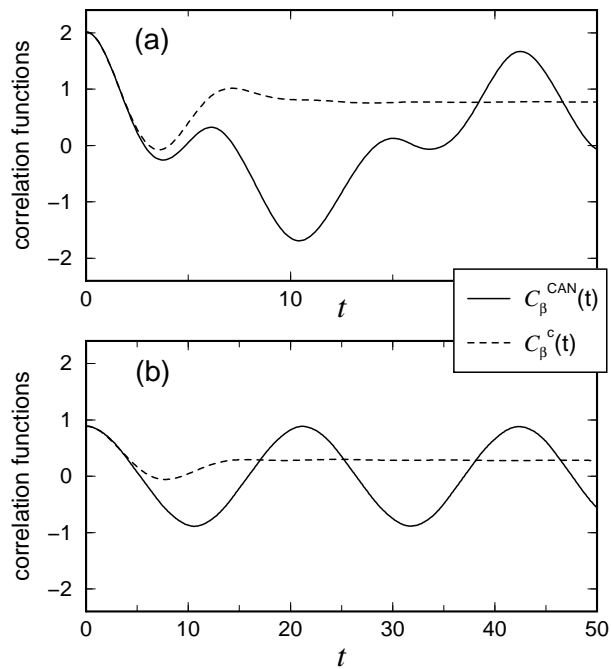}
\vspace{10mm}
\caption{
The exact canonical correlation function $C_{\beta}^{\rm CAN}(t)$ 
and the CMD correlation function $C_{\beta}^{c}(t)$:
(a) for $\beta=1$ and (b) for $\beta=10$.
}
\label{fig:CANc}
\end{figure}

Figure \ref{fig:CANc} shows  
the exact canonical correlation function $C_{\beta}^{\rm CAN}(t)$ 
and the centroid correlation functions $C_{\beta}^{c}(t)$ 
at two temperatures, $\beta=1$ and $10$.
In both cases, we can see that the CMD approximation well 
reproduces the exact short time behavior, as was already
found in Refs. \onlinecite{jv} and \onlinecite{kb}.  
This is because the CMD approximation Eq. (\ref{36}) 
is exact at $t=0$ (see Appendix A). 
However, at $t>0$ the relation Eq. (\ref{36}) does not exactly hold, 
so that the long time CMD behavior 
eventually deviates from the exact one, 
even though the centroid force $F_{\beta}^{c}(q_{c})$ 
is calculated correctly.
For $\beta=10$, the coincidence of $C_{\beta}^{c}(t)$ 
with the exact $C_{\beta}^{\rm CAN}(t)$ persists 
up to $t\simeq 4$, whereas it breaks down at $t\simeq 2.5$
for $\beta=1$.
Namely, the CMD result for $\beta=10$ is better than for $\beta=1$. 
This is because, 
for such shallow double well potential as we used here,  
the shape of $V_{\beta}^{c}(q_{c})$ changes into quasi-harmonic one
as the temperature lowers (see Fig. \ref{fig:ecp});        
this results in a weaker ensemble dephasing effect \cite{jv,kb}. 
However, this property does not hold for deeper double well potentials
whose effective potentials have strongly anharmonic shape \cite{nprg,za}.
\par
\subsection{Effective potential analytic continuation method}
\begin{figure}
\includegraphics[width=80mm,height=80mm]{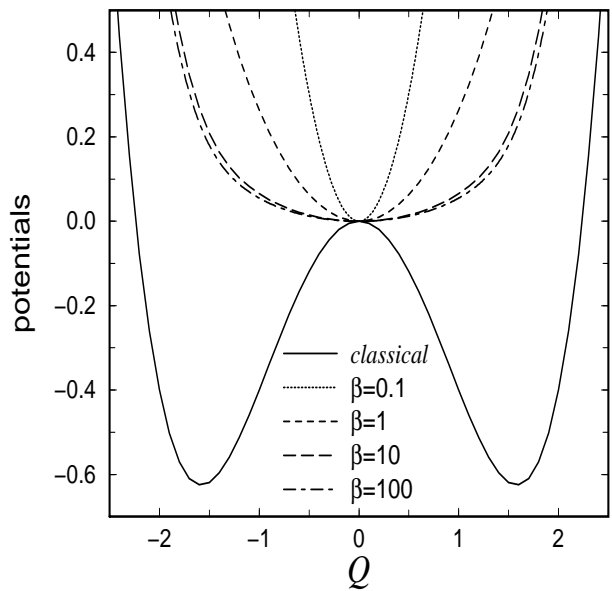}
\vspace{0mm}
\caption{
$\beta$-dependence of the standard effective potential 
$V_{\beta}(Q)$. We set $V_{\beta}(0)=0$.
}
\label{fig:ep}
\end{figure}

\hspace*{\parindent}
Following the procedure presented in Sec. III B,
we have calculated the standard effective potential
$V_{\beta}(Q)$ from the effective classical potential 
$V_{\beta}^{c}(q_{c})$ which was obtained 
from the PIMD calculations in the last subsection. 
Figure \ref{fig:ep} shows 
$V_{\beta}(Q)$ at $\beta=0.1,1,10$ and $100$. 
All of the curves are convex, as analytically
shown in Eq. (\ref{23}).
Especially for $\beta=100$,
$V_{\beta}(Q)$ has almost same 
potential shape as $V_{\beta}^{c}(q_{c})$ in Fig. \ref{fig:ecp}, 
as expected from the relation Eq. (\ref{29}).
As for the high temperature limit,
$V_{\beta}^{c}(q_{c})$ and $V_{\beta}(Q)$ 
have quite different shape.
For $\beta\to 0$, the effective classical potential 
$V_{\beta}^{c}(q_{c})$ is always equal to the classical (bare) potential, 
because there is no quantum correction to $V_{\beta}^{c}(q_{c})$
in the high temperature limit.
On the other hand, the standard effective potential $V_{\beta}(Q)$
generally changes from the classical one even in the high temperature limit, 
because $V_{\beta}(Q)$ receives the corrections 
from the quantum fluctuation and the thermal fluctuation.
In fact, 
for the double well potential considered here,
as suggested in Fig. \ref{fig:ep},
$V_{\beta}(Q)$ in the $\beta\to 0$ limit 
becomes a singular potential which has an infinite   
effective frequency $\omega_{\beta}\to\infty$.  
As an exception, it is well known that for the harmonic oscillator, 
$V_{\beta}^{c}(q_{c})$ and $V_{\beta}(Q)$ have the same 
potential shape at all the temperatures,
because in this case 
the quantum and/or thermal fluctuation
contributes only to the corrections 
to the constant terms in $V_{\beta}^{c}(q_{c})$ and $V_{\beta}(Q)$.
It should also be noted that $Q_{\rm min}$ which gives 
the minimum of $V_{\beta}(Q)$ is always zero
for the classical potential Eq. (\ref{49}) considered here, 
because the potential Eq. (\ref{49}) is 
$Z_{2}$ ($q\leftrightarrow -q$) symmetric and 
$V_{\beta}(Q)$ always has a convex 
($\partial^2 V_{\beta}/\partial Q^2>0$) shape \cite{comment}. 
\par
\par
\begin{figure}
\includegraphics[width=80mm,height=80mm]{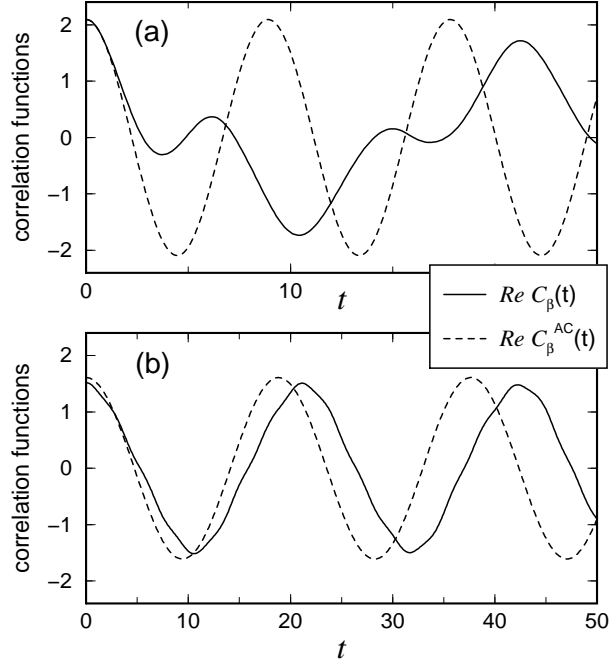}
\vspace{10mm}
\caption{
The real part of the exact quantum correlation function 
$C_{\beta}(t)$ and the real part of the EPAC correlation function 
$C_{\beta}^{\rm AC}(t)$:
(a) for $\beta=1$ and (b) for $\beta=10$.
}
\label{fig:ACc}
\end{figure}

In Fig. \ref{fig:ACc}, we show  
the real part of the exact quantum correlation function 
$C_{\beta}(t)$ 
and the real part of the calculated EPAC correlation function
$C_{\beta}^{\rm AC}(t)$ at $\beta=1$ and $10$.
At the higher temperature [Fig. \ref{fig:ACc}(a)], 
the EPAC method reproduces well
the exact correlation function at short time 
less than $t\sim 2$; 
such a good agreement in the short time behavior 
is similar to the results of the CMD method. 
However, in contrast to the CMD, 
the approximate function $C_{\beta}^{\rm AC}(t)$ 
deviates from the exact quasi-periodic behavior 
at longer time
because of its completely periodic oscillation.    
This is caused by the fact that 
$C_{\beta}^{\rm AC}(t)$ has no dephasing 
because it consists of only one
oscillation mode with the frequency $\omega_{\beta}$
[see Eq. (\ref{46})]. 
On the other hand,
for $\beta=10$, the initial value $C_{\beta}^{\rm AC}(0)$ 
slightly deviates from the exact one, 
while $C_{\beta}^{\rm AC}(t)$ reproduces 
the global oscillating behavior of $C_{\beta}(t)$ 
better than the CMD correlation functions.
This initial value deviation is because  
the EPAC approximation [Eq. (\ref{46})]
is not necessarily exact at $t=0$. 
The initial value $C_{\beta}^{\rm AC}(0)$
can be improved by employing the higher order 
derivative expansion (see the next subsection);
it should become exact in an infinite order expansion.
The EPAC is therefore a method 
to capture the oscillating behavior of quantum systems; 
it should be very effective for systems in which 
quantum coherence is significant.
\par
\par
The essential features of the correlation functions 
can be seen more clearly in the Fourier space.
The Fourier component of the exact real time 
quantum correlation function is 
\begin{eqnarray}
C_{\beta}(\omega)=\int^{\infty}_{-\infty} e^{i\omega t}C_{\beta}(t)
=\frac{1}{{\cal Z}_{\beta}}\sum_{n}\sum_{m}
e^{-\beta \hbar\omega_{n}}(2\pi)~\!\delta(\omega-\omega_{m,n})
\left|\langle m|\hat{q}|n\rangle\right|^{2}
,\label{56a}
\end{eqnarray}
where $\omega_{n}=E_{n}/\hbar$
and $\omega_{m,n}=(E_{m}-E_{n})/\hbar$.
The correlation function $C_{\beta}(\omega)$ 
has a discrete spectrum
and each $\omega_{m,n}$ mode contributes to $C_{\beta}(\omega)$
with the weight $e^{-\beta \hbar\omega_{n}}
\left|\langle m|\hat{q}|n\rangle\right|^{2}$.
Therefore, for finite $\beta$, 
$C_{\beta}(t)$ exhibits quasi-periodic oscillation          
as a result of the interference between 
many $\omega_{m,n}$ modes.
As for the canonical correlation function, 
the Fourier component is 
$C_{\beta}^{\rm CAN}(\omega)=C_{\beta}(\omega)/E(\omega)$,
as expressed in Eq. (\ref{38a}).
Since the function $E(\omega)$ 
is unity at $\omega=0$ and 
is a monotonously increasing function of $\omega$,
the higher frequency modes in $C_{\beta}^{\rm CAN}(\omega)$ 
are less intensive than those in $C_{\beta}(\omega)$.
However, it is noted that the difference 
between $C_{\beta}(\omega)$ and $C_{\beta}^{\rm CAN}(\omega)$
lies just in intensity;
the discrete spectrum coming from the 
delta-function in $C_{\beta}(\omega)$ [Eq. (\ref{56a})]
is also the case of $C_{\beta}^{\rm CAN}(\omega)$.
\par
On the other hand, 
for the CMD correlation function $C^{c}_{\beta}(t)$, 
its Fourier spectrum is, in general, continuous 
because there is no origin of a discrete spectrum 
in the classical definition of  
$C^{c}_{\beta}(t)$ in Eq. (\ref{35}).
Therefore, the thermal fluctuation causes 
the destructive interference 
between an infinite number of  
oscillating modes; this is the ensemble dephasing.
This effect leads to the result that the CMD method underestimates 
the exact quasi-periodic oscillation.
Only the harmonic oscillator is free from 
this dephasing due to the interference.
As for the EPAC correlation function, however, 
its Fourier component has the form
\begin{eqnarray}
C_{\beta}^{\rm AC}(\omega)&=&
\int^{\infty}_{-\infty} e^{i\omega t}C_{\beta}^{\rm AC}(t)
\nonumber\\
&=&
E(\omega)~\!
\frac{\pi}{m\omega_{\beta}^2\beta}\left[~
\delta (\omega+\omega_{\beta})+\delta (\omega-\omega_{\beta})~ 
\right]+(2\pi)~\!\delta(\omega)Q_{\rm min}^{2}
,\label{56b}
\end{eqnarray}
which consists of a single oscillation mode with the 
frequency $\omega_{\beta}$.
Evidently, Eq. (\ref{56b}) gives a discrete spectrum 
as well as $C_{\beta}(\omega)$.
\par
\begin{figure}
\includegraphics[width=80mm,height=80mm]{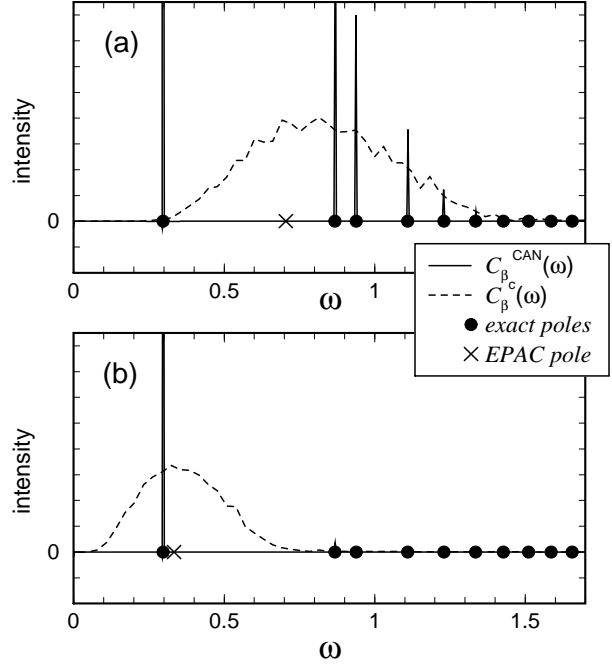}
\vspace{10mm}
\caption{
The Fourier transformed correlation functions,
$C_{\beta}^{\rm CAN}(\omega)$ and  
$C_{\beta}^{c}(\omega)$:
(a) for $\beta=1$ and (b) for $\beta=10$.
The exact poles at $\omega=\omega_{m,n}$ and the EPAC pole 
at $\omega=\omega_{\beta}$ are also plotted.
}
\label{fig:FTc}
\end{figure}

Figure \ref{fig:FTc} shows 
the Fourier transformed correlation functions,
$C_{\beta}^{\rm CAN}(\omega)$ and  
$C_{\beta}^{c}(\omega)$.
The spectra are plotted after reducing 
artificial Fourier ripple due to the finite time cutoff.
The exact poles at $\omega=\omega_{m,n}$ and the EPAC pole 
at $\omega=\omega_{\beta}$ are plotted in the same figure.
As for the Fourier transformed 
canonical correlation function $C_{\beta}^{\rm CAN}(\omega)$,
at both temperatures, 
the peaks exist at the locations of 
a finite number of poles $\omega_{m,n}$.
Hence $C_{\beta}^{\rm CAN}(t)$ is dominated 
by a limited number of oscillation modes, 
leading to the quasi-periodic oscillatory behavior.
On the other hand,
the CMD correlation function $C_{\beta}^{c}(\omega)$
has continuous spectrum for each temperature, 
while the EPAC correlation function
has a discrete spectrum with a peak at $\omega=\omega_{\beta}$.
The CMD and the EPAC methods thus make a remarkable contrast
with each other.
For lower temperature $\beta=10$, in Fig. \ref{fig:FTc}(b), 
fewer oscillation modes are dominant  
in $C_{\beta}^{\rm CAN}(\omega)$
and the EPAC pole $\omega_{\beta}$ exists near the 
exact first pole $\omega_{1,0}$,
while the CMD correlation function $C_{\beta}^{c}(\omega)$ 
still has a broad spectrum.
\par
\par
As mentioned above, 
the EPAC is an approximation to extract  
the effective periodic oscillation from 
the exact correlation function.
Therefore, it can be useful for the investigation of 
the systems where quantum coherence is significant. 
On the other hand, the CMD is a method to approximate
the exact spectrum $C_{\beta}^{\rm CAN}(\omega)$ 
by a continuous spectrum, so that
it is not proper to analyze 
the long time behavior of such systems. 
However, it has been found by Krilov and Berne, 
that the CMD gives much better results in a system 
under isobaric conditions, where the exact correlation function 
has a continuous spectrum \cite{kb}.     
They also suggested that the CMD might 
give fairly accurate results in a dissipative system,
because in such a system 
the quantum coherence is dephased through the interaction 
with the dissipative environment    
not to play an important role at longer time.
\subsection{Possible improvement}
\hspace*{\parindent}
The EPAC method can be improved 
by means of the higher order derivative expansion 
of the effective action $\Gamma_{\beta}[Q]$.
For example, the second order derivative expansion 
in Eq. (\ref{20}) is
\begin{eqnarray}
\Gamma_{\beta}[Q]&=&\int^{\beta\hbar}_{0}d\tau\left[
V_{\beta}(Q)+\frac{1}{2}Z_{\beta}(Q)\dot{Q}^{2}\right]
.\label{56}
\end{eqnarray}
Note the difference from Eq. (\ref{39}).
Then we obtain the correlation function
as a result of the {\it second-order EPAC} ~\cite{note},
\begin{eqnarray}
C^{{\rm AC}(2)}_{\beta}(t)&=&
\left(\frac{\hbar}{2Z_{\beta}\omega_{\beta}^{S}}
\coth\frac{\beta\hbar\omega_{\beta}^{S}}{2}\right)\cos\omega_{\beta}^{S}t
-i\left(\frac{\hbar}{2Z_{\beta}\omega_{\beta}^{S}}\right)
\sin\omega_{\beta}^{S}t+Q_{\rm min}^{2}
,\label{57}
\end{eqnarray}
where $Z_{\beta}=Z_{\beta}(Q_{\rm min})$ and
\begin{eqnarray}
\omega_{\beta}^{S}
&=&\sqrt{\frac{1}{Z_{\beta}}\left.\frac{\partial^2 V_{\beta}}
{\partial Q^2}\right|}_{Q=Q_{\rm min}}
.\label{58}
\end{eqnarray}
The second-order EPAC correlation function 
again consists of a single oscillation mode, 
because the second-order derivative expansion 
only introduces the quantum/thermal correction to 
the particle mass, $m\to Z_{\beta}$, 
and never changes the number of the poles.  
In fact, it is expected that the $n(>2)$th order 
derivative expansion 
improves the qualitative behavior of
the EPAC correlation functions
by introducing additional poles.
In such cases, 
the $n$th order EPAC correlation function
$C_{\beta}^{{\rm AC}(n)}(t)$ would exhibit 
quasi-periodic oscillation 
more similar to the exact behavior.
\par
\begin{figure}
\includegraphics[width=80mm,height=80mm]{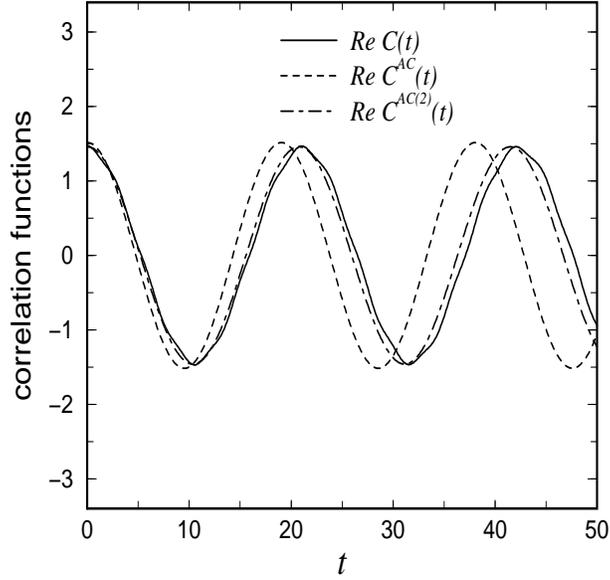}
\vspace{0mm}
\caption{
The real parts of the quantum correlation functions
at zero temperature: 
the exact quantum correlation function $C(t)$, 
the leading order EPAC correlation function $C^{\rm AC}(t)$, 
and 
the second-order EPAC correlation function  $C^{{\rm AC}(2)}(t)$.
}
\label{fig:ACT0}
\end{figure}
\par
To see the effectiveness of this improvement, 
we now present a zero temperature example.
At zero temperature, the exact real time quantum correlation 
function is 
\begin{eqnarray}
C(t)&=&\langle 0|~\!\hat{q}(t)\hat{q}(0)|0\rangle= 
\sum_{m}e^{-i(E_{m}-E_{0})t/\hbar}
\left|\langle m|\hat{q}|0\rangle\right|^{2}
.\label{59}
\end{eqnarray}
On the other hand, the leading order EPAC correlation function is
\begin{eqnarray}
C^{\rm AC}(t)&=&\frac{\hbar}{2 m \omega_{\rm eff}}
e^{-i \omega_{\rm eff}t} + Q^{2}_{\rm min}
,\label{60}
\end{eqnarray}
where $\omega_{\rm eff}=\omega_{\beta}|_{\beta\to \infty}$,
and the second order EPAC correlation function 
obtained from Eq. (\ref{57}) is
\begin{eqnarray}
C^{{\rm AC}(2)}(t)&=&\frac{\hbar}{2 Z_{\rm eff} \omega_{\rm eff}^{S}}
e^{-i \omega_{\rm eff}^{S} t} + Q^{2}_{\rm min}
,\label{61}
\end{eqnarray}
where $Z_{\rm eff}=Z_{\beta}|_{\beta\to \infty}$ and 
$\omega_{\rm eff}^{S}=\omega_{\beta}^{S}|_{\beta\to \infty}$.
To evaluate these zero temperature quantities, 
$\omega_{\rm eff}$, $Z_{\rm eff}$, and $\omega_{\rm eff}^{S}$, 
we employed 
the renormalization group technique \cite{wk} for convenience; 
the effective frequency $\omega_{\rm eff}$ 
was calculated by solving the
local potential approximated 
Wegner-Houghton equation \cite{wh,nprg},
while for $Z_{\rm eff}$ and $\omega_{\rm eff}^{S}$, 
we used the values obtained by solving 
the proper time renormalization group equation \cite{za}. 
Figure \ref{fig:ACT0} shows  
the real part of the exact quantum correlation function 
$C(t)$ 
and the real parts of the EPAC correlation functions: 
the leading order $C^{\rm AC}(t)$ and 
the second order $C^{{\rm AC}(2)}(t)$. 
It is seen that the deviation of the leading order EPAC
correlation function $C^{\rm AC}(t)$  
from the exact quantum correlation function  
is improved in the second-order EPAC ($C^{{\rm AC}(2)}(t)$),
both in its initial ($t=0$) value and 
in its long time behavior.
\par 
The higher order derivative expansion is required 
especially for the analyses of many-body systems at higher temperature,
because the multipole contribution to the correlation functions
is significant for describing the decay of them.
However, it is not easy to compute the higher derivative terms 
in $\Gamma_{\beta}$, especially at finite temperature.   
Therefore, it is a future task 
to establish an algorithm for computing them.  
\section{CONCLUDING REMARKS}
\hspace*{\parindent}
As a novel approximation method to evaluate 
the real time quantum correlation 
functions at finite temperature,
we have newly proposed the EPAC method.
The EPAC method has been tested in the
one-dimensional symmetric double-well system,
in comparison with another approximation scheme, the CMD method.
The EPAC method and the CMD method
are based on each different type of effective potential,
the standard effective potential $V_{\beta}(Q)$ and 
the effective classical potential $V_{\beta}^{c}(q_{c})$,
respectively.
At first, we have evaluated $V_{\beta}^{c}(q_{c})$ by means of  
the normal mode PIMD calculation,
and then $V_{\beta}(Q)$ have been obtained from it.
It has been found that 
these effective potentials 
have the predicted formal properties:
The equivalence of them in the zero temperature limit and 
the convexity of $V_{\beta}(Q)$.
Then the real time two-point position correlation functions
have been calculated 
by means of each approximation method.
The CMD approximation is found to be good at short time range 
owing to its exactness at $t=0$, while at longer time 
it largely deviates 
from the exact correlation function 
because of the ensemble dephasing. 
On the other hand, our EPAC approximation 
can reproduce the long time oscillating behavior
which originates from the quantum coherence of the system. 
Therefore, the EPAC should be very effective for the system in which 
quantum coherence is significant,
such as the quantum double well system considered here.
We have also suggested that the EPAC method can be improved 
by the higher order derivative expansion, 
and have shown, as an example, 
the result of the second order improvement 
at zero temperature.
It has been seen that 
this improvement procedure works very well
for this example.
\par
In this paper we have restricted the arguments
to the evaluation of 
the real time two-point position correlation function
$C_{\beta}(t)=\langle \hat{q}(t)\hat{q}(0)\rangle_{\beta}$.
Also for general correlation functions of 
nonlinear operators, e.g.,
$\langle \hat{q}^n(t)\hat{q}^n(0)\rangle_{\beta}$,
the EPAC can be applied to evaluate them, 
because the $n$-point correlation function is obtained 
with the knowledge of the $n$th functional derivative 
of the effective action \cite{riv,swa,ps,kl}.
This will be an interesting subject to be examined 
in near future.
\par
As matters of theoretical interest, 
there are a number of subjects to be investigated.
The role of the 
standard effective potential 
in the quantum transition-state theory \cite{gil,vm}
should be clarified. 
Furthermore, it is interesting to find 
possible relationship between 
the effective potential-based methods 
of the present type and the other approaches 
to quantum dynamical correlations \cite{blt}.
It is worthwhile to pay attention to the fact that, 
for example, the effective potential-based quantum dynamics  
has been discussed in the context of 
the projection operator approach \cite{cgtv}.
\par
Finally we mention the direction of the applications 
of the EPAC to real molecular systems of chemical interest.
From the results given in Sec. IV, 
it is suggested that  
the EPAC should be very effective 
for the systems where quantum coherence plays an important role.
For example, proton transfer reactions \cite{proton}
are typical phenomena of such category, 
because the small mass of proton makes  
quantum coherence significant even at room temperature. 
The EPAC will properly represent 
the quantum oscillating behavior 
accompanied with proton transfers. 
For example, for a molecular reaction system where 
the intrinsic reaction coordinate (IRC) \cite{fukui}
is well-defined,
once the IRC and the potential energy surface 
along it are provided, 
it is straightforward to calculate the approximate 
real time quantum correlation function 
by use of the EPAC method.
However, for reactions occurring in solvents,
not only the IRC but the full quantum calculations
treating many degrees of freedom 
should be implemented.
For the application of the EPAC 
to such many-body systems including quantum coherence, 
more efficient sampling algorithms 
or novel approximation schemes
must be developed  
to calculate the effective classical potentials
needed in the EPAC.
\begin{acknowledgments}
This work was supported by a fund for Research and Development 
for Applying Advanced Computational Science and Technology,
Japan Science and Technology Corporation (ACT-JST).
\end{acknowledgments}
\appendix
\section{ON THE RELATION BETWEEN 
$C^{c}_{\beta}(0)$ and $C^{\rm CAN}_{\beta}(0)$}
\hspace*{\parindent}
It has been known that the CMD approximation [Eq. (\ref{36})] 
is exact at $t=0$, i.e., $C^{c}_{\beta}(0)=C^{\rm CAN}_{\beta}(0)$ 
\cite{jv,rrjv,ra}.
Note that this relation holds only for 
the correlation function of {\it linear} operator,
as seen in the following simple derivation:    
\begin{eqnarray}
C_{\beta}^{\rm CAN}(0)&=&\frac{1}{\beta}\int_{0}^{\beta}d\lambda
~\langle \hat{q}(-i\hbar\lambda)\hat{q}\rangle_{\beta}
\nonumber\\
&=&\frac{1}{\beta\hbar}\int_{0}^{\beta\hbar}d u
~\langle \hat{q}(-iu)\hat{q}\rangle_{\beta}
\nonumber\\
&=&\langle \hat{q}_{0}\hat{q}\rangle_{\beta}
\nonumber\\
&=&\langle \hat{q}_{0}~\!
\frac{1}{\beta\hbar}\sum_{n=-\infty}^{\infty}
\hat{q}(\omega_n)\rangle_{\beta}
\nonumber\\
&=&\langle \hat{q}_{0}~\!\hat{q}_{0}\rangle_{\beta}
\nonumber\\
&=&\frac{1}{{\cal Z}_{\beta}}\int^{\infty}_{-\infty}dq
\int^{q(\beta\hbar)=q}_{q(0)=q}{\cal D}q
~\!q_{0}~\!q_{0}~\!e^{-S_{E}/\hbar}
\nonumber\\
&=&\frac{1}{{\cal Z}_{\beta}}
\int^{\infty}_{-\infty}dq_{c}\int^{\infty}_{-\infty}dq
\int^{q(\beta\hbar)=q}_{q(0)=q}{\cal D}q
~\!q_{0}~\!q_{0}~\!\delta(q_{0}-q_{c})~\!e^{-S_{E}/\hbar}
\nonumber\\
&=&\frac{1}{{\cal Z}_{\beta}}
\int^{\infty}_{-\infty}dq_{c}~\!q_{c}~\!q_{c}~\!
\int^{\infty}_{-\infty}dq
\int^{q(\beta\hbar)=q}_{q(0)=q}{\cal D}q
~\!\delta(q_{0}-q_{c})~\!e^{-S_{E}/\hbar}
\nonumber\\
&=&\frac{1}{{\cal Z}_{\beta}}
\int^{\infty}_{-\infty}dq_{c}~\!q_{c}~\!q_{c}~\!
\rho_{\beta}^{c}(q_{c})
\nonumber\\
&=&\langle q_{c}~\!q_{c}\rangle_{\rho_{\beta}^{c}}
\nonumber\\
&=&C_{\beta}^{c}(0)
,\label{a1}
\end{eqnarray}
where we have used 
$q(\tau)=(1/\beta\hbar)\sum_{n=-\infty}^{\infty}
e^{-i\omega_{n}\tau}q(\omega_n)$,
\begin{eqnarray}
\langle \hat{q}(-\omega_n)~\!\hat{q}(\omega_m)\rangle_{\beta}
&=&\delta_{nm}\frac{\beta\hbar}{2}\int^{\beta\hbar}_{-\beta\hbar}
d u e^{-i\omega_{n} u}\langle \hat{q}(u)~\!\hat{q}(0)\rangle_{\beta}
,\label{a2}
\end{eqnarray}
and the definition of path centroid $q_{0}=q(\omega_{0})/\beta\hbar$. 


\begin{thebibliography}{99}
\bibitem{eht}
It should be also noted that the correlation function 
$C_{\rm ref}(t)$
in the analytically continued effective harmonic theory 
is obtained from the superposition of 
the correlation functions
for the effective harmonic oscillators 
defined at each centroid position $q_{c}$ 
with frequency $\bar{\omega}_{c}(q_{c})$.
Each correlation function is then weighted 
by the centroid density $\rho_{\beta}^{c}(q_{c})$, 
and therefore $C_{\rm ref}(t)$ exhibits the ensemble dephasing 
like the CMD correlation function $C_{\beta}^{c}(t)$.
\bibitem{comment}
For an asymmetric potential, $Q_{\rm min}$ has a finite value 
and the EPAC correlation function is shifted by 
a constant term $Q_{\rm min}^{2}$ according to Eq. (\ref{46}).
\end{thebibliography}
\end{document}